\documentclass[a4paper,11pt]{article}
\usepackage{pos}

\usepackage{amsmath}
\usepackage{graphicx}
\usepackage{xcolor}
\usepackage{bm}
\usepackage{algorithm}
\usepackage{algpseudocode}
\usepackage{soul}

\title{Chiral rank-$k$ truncations for the multigrid preconditioner of Wilson fermions in lattice QCD}
\ShortTitle{Chiral rank-$k$ truncations for multigrid in LQCD}

\author*[a]{Travis Whyte}
\author[b]{Andreas Stathopoulos}
\author[c]{Eloy Romero}

\affiliation[a]{J\"ulich Supercomputing Center,\\
  J\"ulich, Germany}

\affiliation[b]{Department of Computer Science, William \& Mary,\\
Williamsburg, VA, USA}

\affiliation[c]{Thomas Jefferson National Accelerator Facility, \\
Newport News, VA, USA}

\emailAdd{t.whyte@fz-juelich.de}

\abstract{We present a modification to the setup algorithm for the multigrid preconditioner of Wilson fermions in lattice QCD. A larger number of test vectors than that used in conventional multigrid is generated by the smoother. This set of test vectors is then truncated by a singular value decomposition on the chiral components of the test vectors, which are subsequently used to form the prolongation and restriction matrices of the multigrid hierarchy. This modification is demonstrated to improve the convergence of linear equations on an anisotropic lattice with $m_{\pi} \approx 239$ MeV from the Hadron Spectrum Collaboration and an isotropic lattice with $m_{\pi} \approx 220$ MeV from the MILC Collaboration. The lattice volume dependence of the method is also examined.}

\FullConference{The 41st International Symposium on Lattice Field Theory (LATTICE2024)\\
 28 July - 3 August 2024\\
Liverpool, UK\\}

\begin{document}
\maketitle

\section{Introduction}
The most efficient and widely used method of solving linear equations for the generation of propagators in lattice QCD is multigrid ~\cite{wilsonmg,domainwallmg,staggmg,overlapmg,ddalphamg,twistedmassmg,brannick_schwinger,whyte_schwinger,morgan_twogrid}. The multigrid preconditioner is created by exposing the near null space of the Wilson-Dirac operator, $\bm{D}$, through the generation of test vectors. These test vectors are rich in components that correspond to the low eigenmodes of $\bm{D}$ and are used to create the coarse grid matrices defining the multigrid preconditioner hierarchy. The number of test vectors used to create the coarse grid matrices must be chosen judiciously. Too many test vectors results in larger and denser coarse grid matrices which are expensive to apply. Too few test vectors will result in a preconditioner which cannot effectively reduce the error, leading to an increase in the number of iterations required of the outer iterative solver. 

In this article, a modification of the setup method for the multigrid preconditioner of Wilson type fermions is presented, which improves the basis of test vectors used to create the coarse grid matrices by calculating an enlarged basis of test vectors. This basis is subsequently truncated using a singular value decomposition, which retains the components of the test vectors that have the largest contribution to the basis. Numerical results are reported for anisotropic lattices from the Hadron Spectrum Collaboration and isotropic lattices from the MILC Collaboration.

\section{Multigrid for Wilson Fermions}

In the adaptive multigrid framework, the smoother is applied on the homogenous equation
\begin{equation}
	\bm{D}\psi_i \approx 0, ~~ ~~i = 1, \dots, m
\end{equation}
with $m$ random initial guesses to create a set of $m$ near null or test vectors, $\bm{\Psi} = [\psi_1,\dots,\psi_m]$~\cite{adapt}. This reveals the components of the error that the smoother is not able to easily resolve, e.g. the low frequency components of the error. These smooth test vectors are then blocked according to a domain decomposition of the underlying lattice, which then become the columns of the \textit{prolongator} matrix, $\bm{P}$. The \textit{restriction} matrix, $\bm{R}$, is typically taken to be $\bm{R} = \bm{P}^{\dag}$. The test vectors generated display the phenomenon known as \textit{local coherence}, wherein the smooth modes approximate the low eigenspace of $\bm{D}$ on a given local domain~\cite{luscher_deflation,luscher_coherence}. However, the success of the multigrid preconditioner for the Wilson-Dirac operator relies on ``chiral splitting'', wherein the smooth test vectors are split into chiral components using the projectors $1 \pm \gamma_5$ \cite{wilsonmg}. The application of these projectors to the smooth test vectors splits the spin components of $\bm{\Psi}$ such that
\begin{align}
\label{eq::chiral_split}
	\bm{\Psi}^{(\alpha = 0)} & = [\psi_1^{(\beta = 1,2)},\dots,\psi_m^{(\beta = 1,2)}] \\
	\bm{\Psi}^{(\alpha = 1)} & = [\psi_1^{(\beta = 3,4)},\dots,\psi_m^{(\beta = 3,4)}] , \nonumber
\end{align}
where $\alpha$ labels the chiral index and $\beta$ labels the spin index of the test vectors. The chirally split test vectors are then orthonormalized on a given domain of the spacetime lattice $\Lambda_j \in \Lambda = \{\Lambda_1,\dots,\Lambda_d\}$ and the prolongator is formed as
    \begin{equation}
	    P_{\alpha ij}(x) = \begin{cases} 
      \psi^{\alpha}_i(x) & \mathrm{if}~x\in \Lambda_j \\
      0 & \mathrm{otherwise}
   \end{cases},
    \label{eq:prolong}
    \end{equation}
where $x$ denotes the lattice coordinate and color indices have been suppressed. This results in a block diagonal matrix of dimension $n \times 2md$, where $d = |\Lambda|$. The use of the projectors $1 \pm \gamma_5$ enforces the symmetry $\gamma_5 \bm{P} = \bm{P} \sigma_3$, where $\sigma_3$ is the third Pauli spin matrix, which ensures that if a right singular vector is in the range of $\bm{P}$, a left singular vector is in the range of $\bm{R}$. The coarse grid matrix is formed by
\begin{equation}
	\bm{D}_{\ell+1} = \bm{RD}_{\ell}\bm{P}.
\end{equation}
This process is repeated recursively beginning with the coarse grid operator $\bm{D}_{\ell+1}$ to form successively coarser grids. The projector $(1\pm \gamma_5^{\ell})$ is used to perform the chiral splitting, where $\gamma_5^{\ell} = \sigma_3$ for $\ell > 0$. Typical applications of multigrid in lattice QCD utilize the multigrid hierarchy in a $K$-cycle~\cite{kcycl} as a preconditioner for a flexible outer solver such as FGMRES or GCR.

\section{Chiral Rank-$k$ Truncations of the Multigrid Basis}

The use of rank-$k$ truncations for the multigrid basis of smooth test vectors was given in Ref.~\cite{chow} and was shown to increase the performance of the multigrid preconditioner for two-dimensional anisotropic heat diffusion equations. It was also used in Ref.~\cite{dambra} to form composite aggregate preconditioners in bootstrap AMG. The extension of this method to Wilson fermions in lattice QCD is straightforward. Guided by the required chiral splitting in the conventional set up method, the singular value decomposition is performed separately on the chiral components of the initial set of $m$ smooth test vectors, with $m \geq k$, given by Eq. (\ref{eq::chiral_split}). Let $\bm{U}\bm{\Sigma}\bm{V}^{\dag} = \bm{\Psi}^{(\alpha)}(\Lambda_j)$ be the singular value decomposition of $\bm{\Psi}^{(\alpha)}(\Lambda_j)$. Then
\begin{equation}
	\bm{U}^{(\alpha)}_k\bm{\Sigma}^{(\alpha)}_k\bm{V}_k^{(\alpha)\dag} \approx \bm{P}_{\alpha j}\bm{\Psi}^{(\alpha,\ell+1)}(x_c)
	\label{eq::chiral_approx}
\end{equation}
with $\alpha = 0, 1$, is the best rank-$k$ approximation to the initial basis $\bm{\Psi}^{(\alpha)}(\Lambda_j)$. By matching variables in Eq.~(\ref{eq::chiral_approx}), we observe that $\bm{U}^{(\alpha)}_k = \bm{P}_{\alpha j}$ and $\bm{\Sigma}^{(\alpha)}_k\bm{V}_k^{(\alpha)\dag} = \bm{\Psi}^{(\alpha,\ell+1)}(x_c) $, with $x_c$ a coarse grid coordinate. 
The prolongator is then formed as
    \begin{equation}
            P_{\alpha ij}(x) = \begin{cases}
		    u^{(\alpha)}_i(x) & \mathrm{if}~x\in \Lambda_j \\
      0 & \mathrm{otherwise}
   \end{cases}
    \label{eq::chiral_prolong}
    \end{equation}
 where $\bm{P}$ is of dimension $n \times 2kd$. As the singular value decomposition creates a basis of $m$ orthonormal left singular vectors, the orthonormalization step of the conventional method on the truncated basis is not needed.

\section{Numerical Experiments}
The setup method with chiral rank-$k$ truncations is tested against conventional multigrid for two different types of lattice ensembles. The first, which we refer to as Ensemble A, corresponds to an anisotropic lattice of dimension $32^3 \times 256$ with a pion of mass $m_{\pi} \approx 239$ MeV from the Hadron Spectrum Collaboration~\cite{hs1,hs2}. The gauge links arising from these configurations are stout smeared~\cite{stout}. See Ref.~\cite{hs3} for more information relating to these lattices, including scale and mass determination. The second, which we refer to as Ensemble B, corresponds to an isotropic lattice of dimension $32^3 \times 64$ with a pion of mass $m_\pi \approx 220$ MeV from the MILC Collaboration~\cite{milc1,milc2}. In this Clover on HISQ action, the gauge links are smeared with one iteration of HYP smearing~\cite{hyp}. Ref.~\cite{isovector_charges} contains more information on the determination of the relevant physical properties of this ensemble. All experiments were performed on eight nodes with Intel Xeon Gold 6130 processors with $2 \times 16$ cores. In all experiments, a hierarchy of three multigrid operators is used and the results are averaged over ten right hand sides.
\label{sec:exp}

\subsection{Size of the Multigrid Basis}
\label{subsec:size}
As the size of the initial multigrid basis $m$ is increased, it is natural to expect that more information about the near null space is captured by the preconditioner, leading to improved convergence of the linear equations. Typical applications of multigrid in lattice QCD use between 24 and 32 test vectors. However, the chiral rank-$k$ truncation allows us to calculate an initial basis size of $m \geq k$ and incorporate information from the test vectors that the conventional setup method ignores. It is thus important to quantify the performance increase offered by increasing $m$.
\begin{figure}[!h]
	\begin{center}
	\includegraphics[scale=0.425]{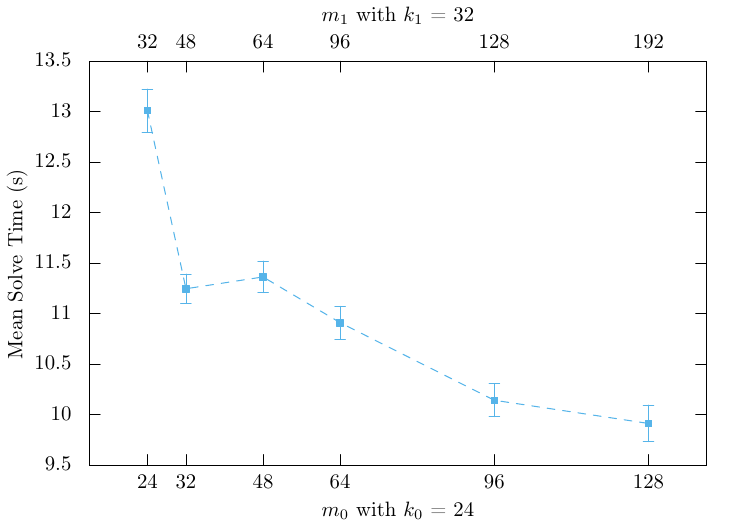}
	\includegraphics[scale=0.425]{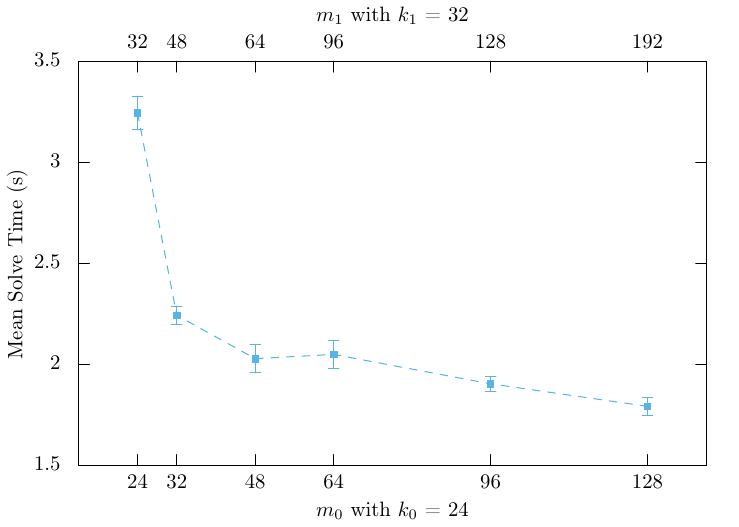}
		\caption{\label{fig::basissize}The mean solve time as a function of the initial basis size for Ensemble A (left) and Ensemble B (right). The bottom and top $y$-axis displays $m$ for levels $\ell = 0,1$, respectively.}
	\end{center}
\end{figure}
Fig.~\ref{fig::basissize} shows the mean execution time for solving the systems of linear equations for both Ensemble A and Ensemble B. The basis is truncated at $k = 24,32$ for levels $\ell = 0,1$, respectively. A general trend is observed for both ensembles: as $m$ is increased, the mean execution time decreases. This is inline with expectations that additional information from the exposed smooth modes is made available to the preconditioner. It is also observed that a saturation in the speedup begins between $m = 96$ and $m = 128$, indicating that the modes beyond $m = 96$ contribute very little to reducing the error.
\subsection{Optimal Truncation}
\label{subsec:opt_trunc}
The number of test vectors used to create the prolongation and restriction matrices is an important determination. Too few vectors results in a poor preconditioner and too many test vectors results in large and dense coarse grid matrices that are expensive to apply. It is thus important to examine the rank of the truncation utilized.  
\begin{figure}[!h]
	\begin{center}
	\includegraphics[scale=0.425]{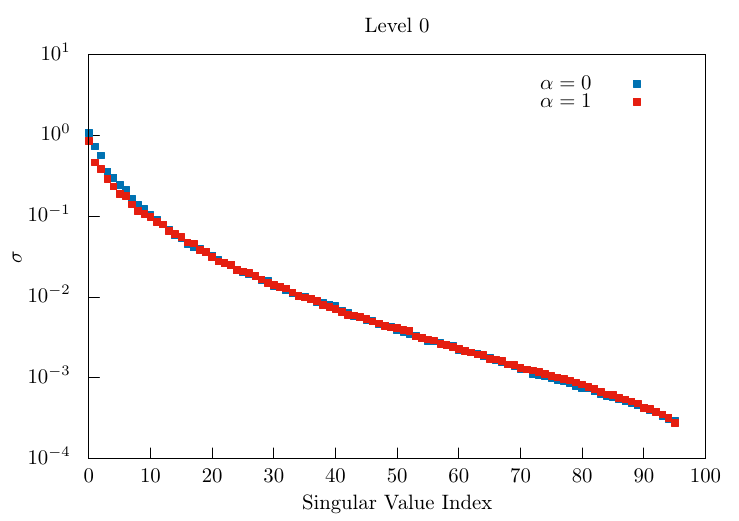}
	\includegraphics[scale=0.425]{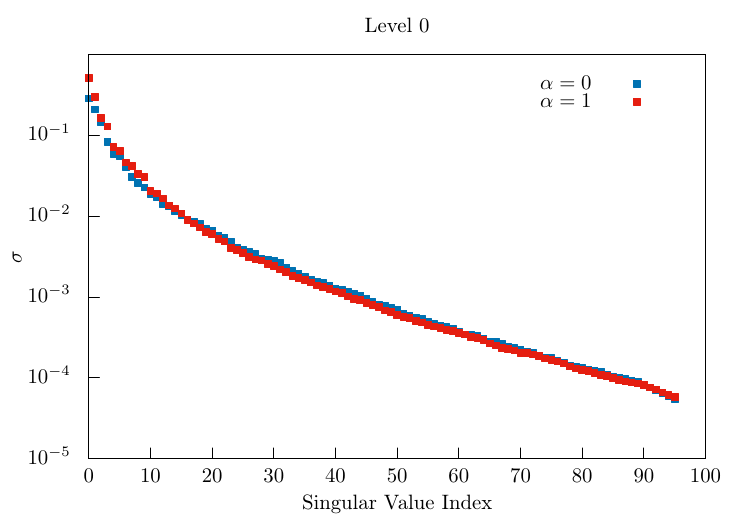}
	\includegraphics[scale=0.425]{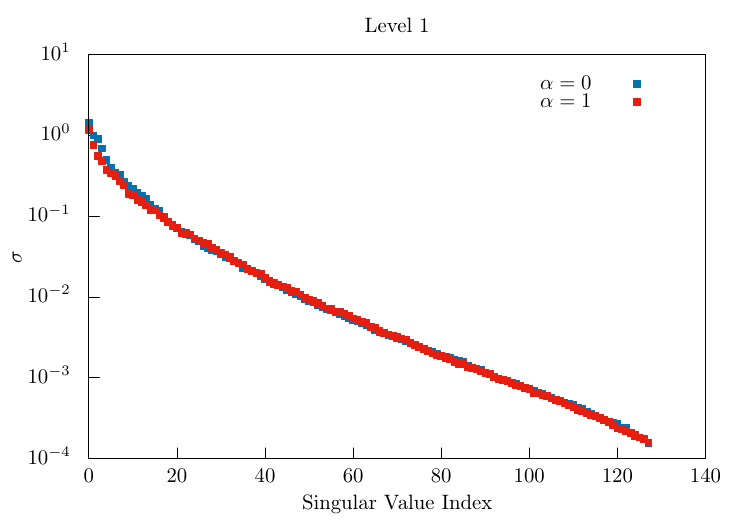}
	\includegraphics[scale=0.425]{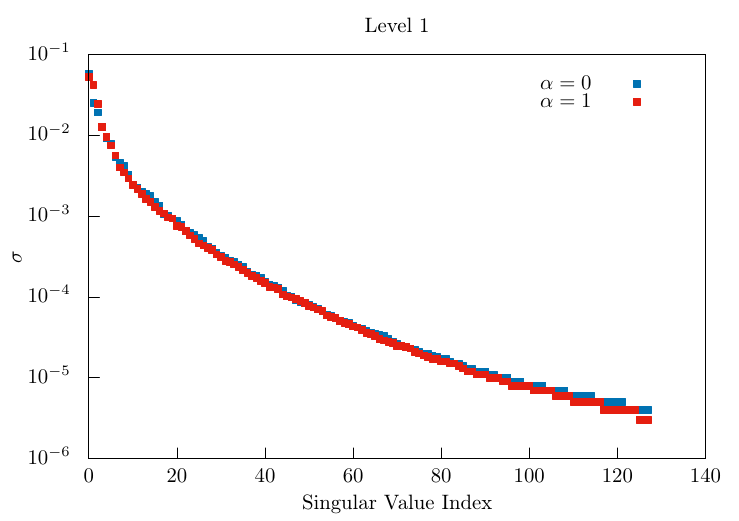}
		\caption{\label{fig::svs}The singular spectrum for the chirally split test vectors on the first domain for level $\ell = 0$ (top) and $\ell = 1$ (bottom) of Ensemble A (left) and Ensemble B (right).}
	\end{center}
\end{figure}
The singular spectra of the test vectors provides some information for this determination. Fig.~\ref{fig::svs} shows the singular spectrum of the chirally split test vectors for the first domain, which is representative of other domains. It is observed that the spectrum for both ensembles displays a rapid decay of the singular values. The magnitude of the 24th singular value is approximately two orders of magnitude lower than that of the first. This indicates that including left singular vectors into the basis beyond this approximate cutoff will have a small overall contribution to the low rank approximation of the multigrid basis. 

\begin{figure}[!h]
	\begin{center}
	\includegraphics[scale=0.45]{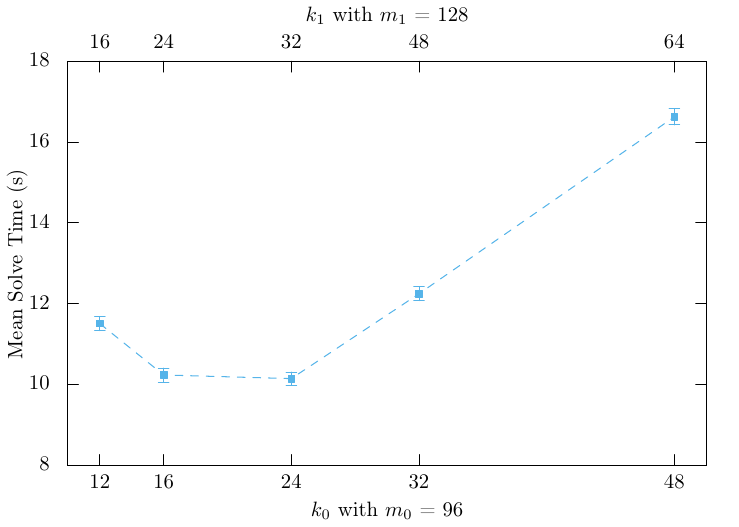}
	\includegraphics[scale=0.45]{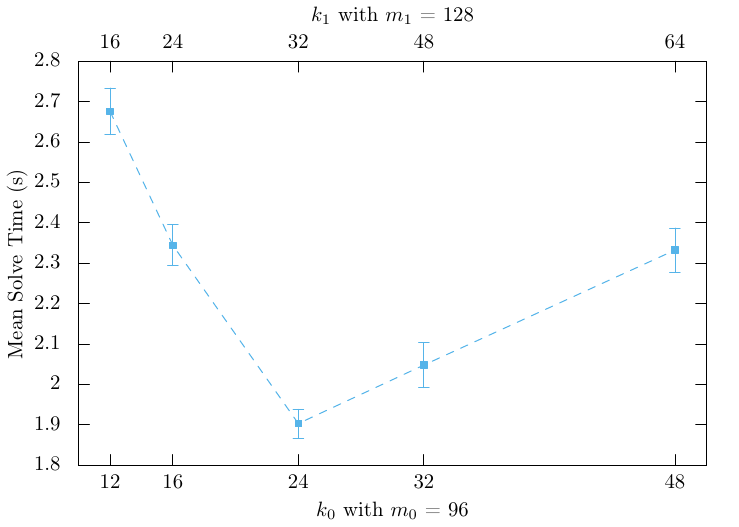}
		\caption{\label{fig::rank}The mean execution time for the system of linear equations when the degree of the truncation is varied using an initial basis size of $m = 96$ for Ensemble A (left) and Ensemble B (right). The bottom and top $y$-axis displays $k$ for levels $\ell = 0,1$, respectively.}
	\end{center}
\end{figure}

Fig.~\ref{fig::rank} displays the mean execution time of the system of linear equations when the degree of the truncation, $k$, is varied for a constant value of $m = 96$. This value of $m$ was chosen as larger values of $m$ provide only a small decrease in execution time. For both ensembles, the optimal truncation is observed to be at $k = 24$. For Ensemble A this is comparable to $k = 16$, while Ensemble B exhibits a distinct minimum at $k = 24$. 

\subsection{Optimal Number of Setup Iterations}
\label{subsec:setup}
The number of set up iterations of the smoother to generate the test vectors may also have a large effect on the efficacy of the preconditioner. It is thus beneficial to examine the performance of the preconditioner while varying the number of setup iterations while generating the test vectors. 
\begin{figure}[!h]
	\begin{center}
        \includegraphics[scale=0.45]{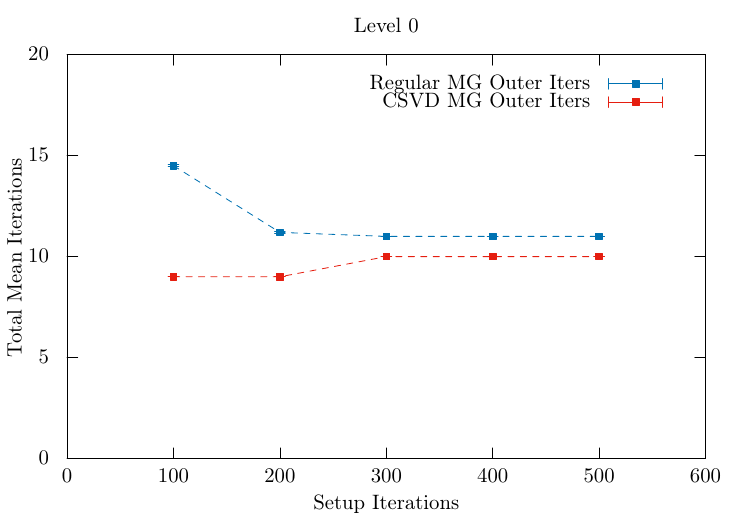}
        \includegraphics[scale=0.45]{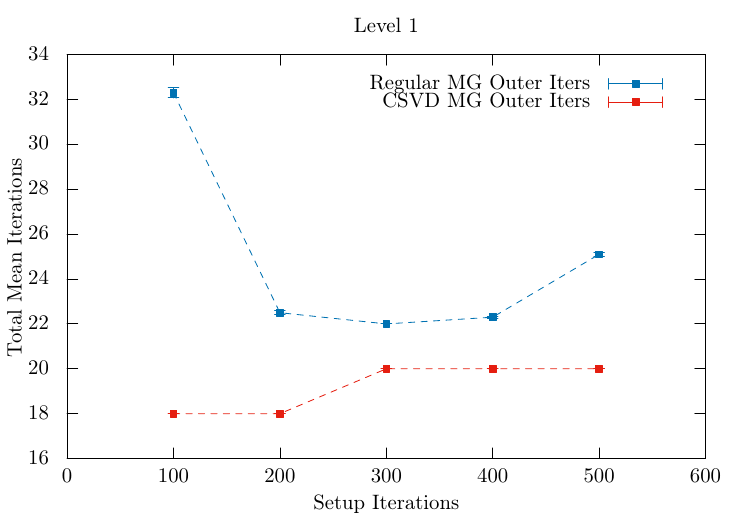}
        \includegraphics[scale=0.45]{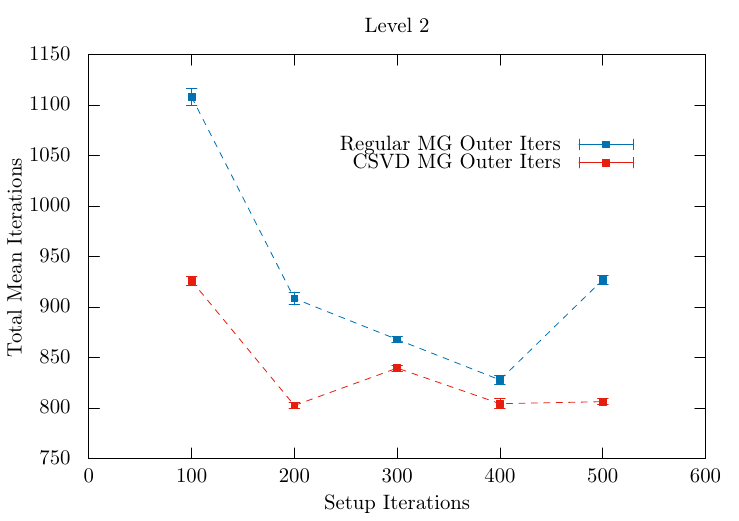}
        \includegraphics[scale=0.45]{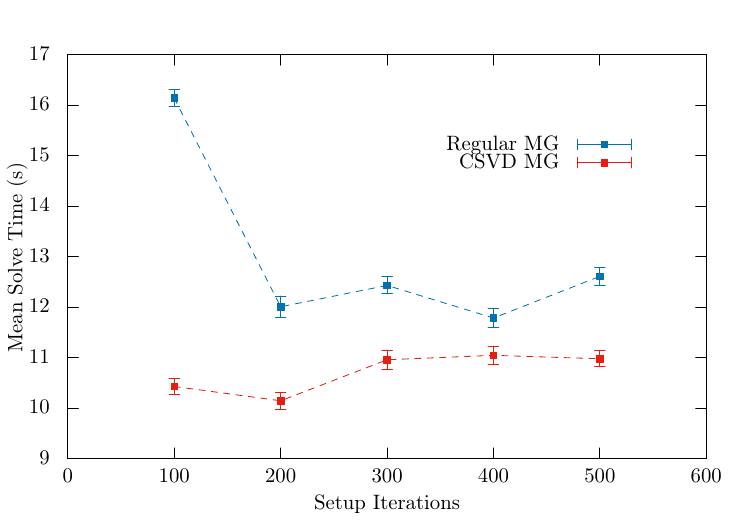}
	\caption{\label{fig::iters_aniso}The total number of iterations on $\ell = 0$ (upper left), $\ell = 1$ (upper right), $\ell = 2$ (lower left) and the mean solve time of the system of linear equations for Ensemble A.}
	\end{center}
\end{figure}

\begin{figure}[!h]
	\begin{center}
        \includegraphics[scale=0.45]{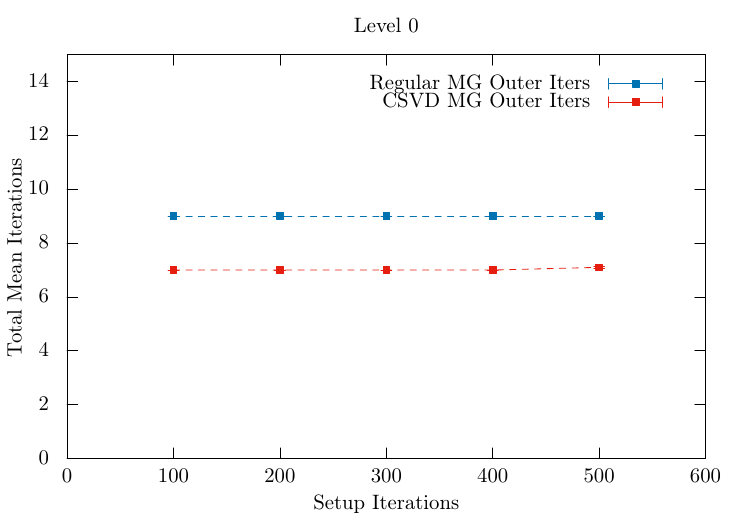}
        \includegraphics[scale=0.45]{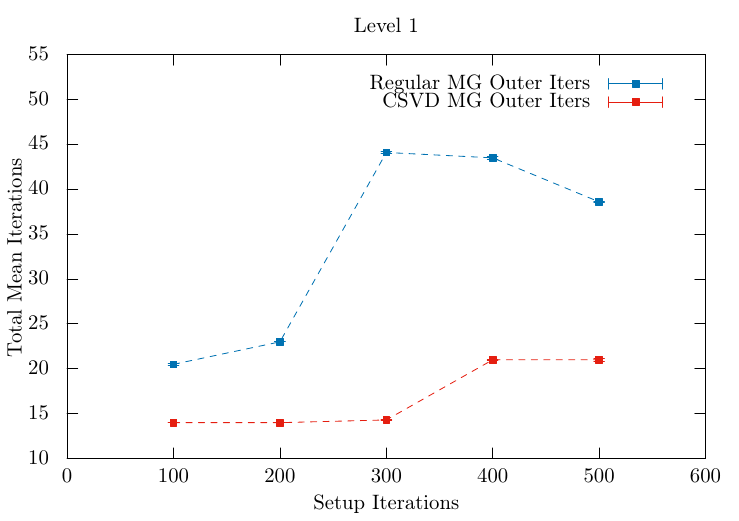}
        \includegraphics[scale=0.45]{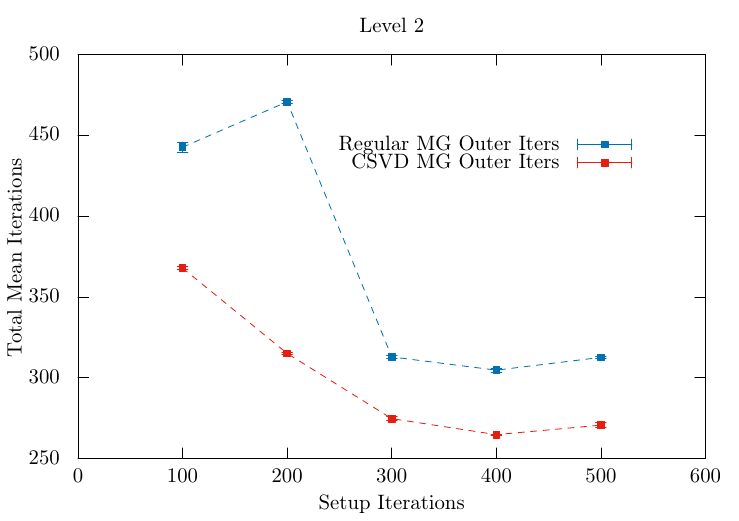}
        \includegraphics[scale=0.45]{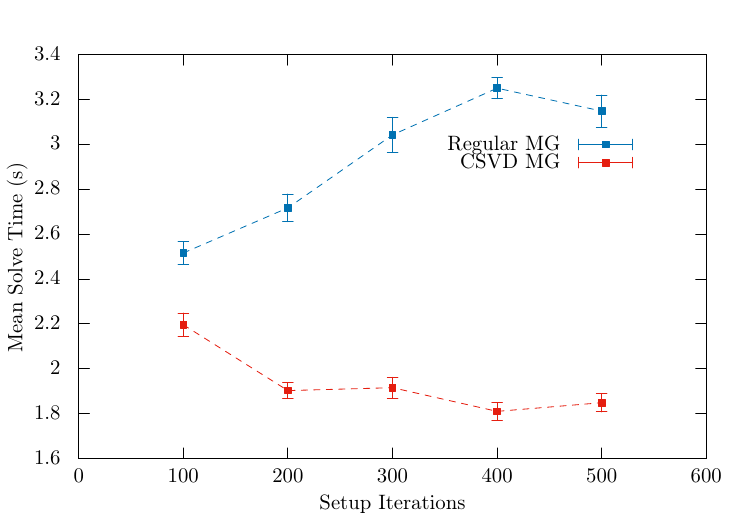}
	\caption{\label{fig::iters_milc}As Fig.~\ref{fig::iters_aniso} for Ensemble B.}
	\end{center}
\end{figure}

Figs. \ref{fig::iters_aniso} and \ref{fig::iters_milc} display the total number of iterations required on each level $\ell$ as well as the mean execution time. For both ensembles, a value of $m = 96$ and $k = 24$ is used, which was found to be the optimal values as observed in Figs. \ref{fig::basissize} and \ref{fig::rank}.  In both cases, it is observed that the total cost of solving the system of linear equations is reduced in comparison to conventional multigrid for all number of setup iterations on every level. Additionally, the use of the chiral rank-$k$ truncation results in a preconditioner that is less sensitive to the number of setup iterations.

\begin{figure}[!h]
	\begin{center}
        \includegraphics[scale=0.45]{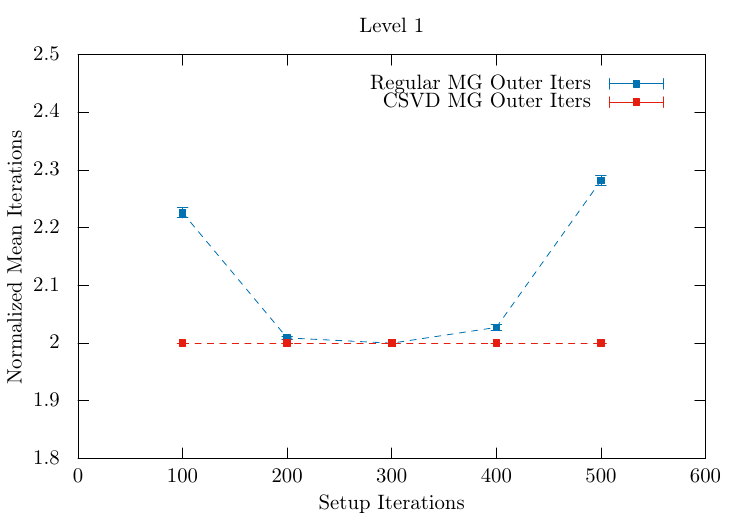}
	\includegraphics[scale=0.45]{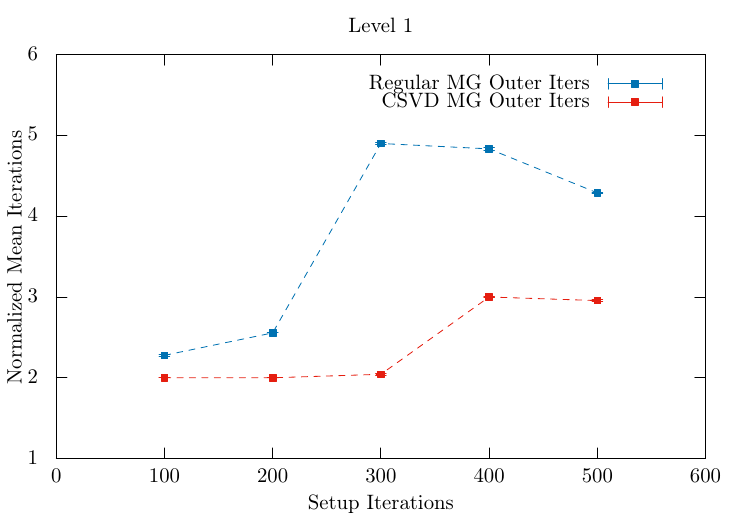}
        \includegraphics[scale=0.45]{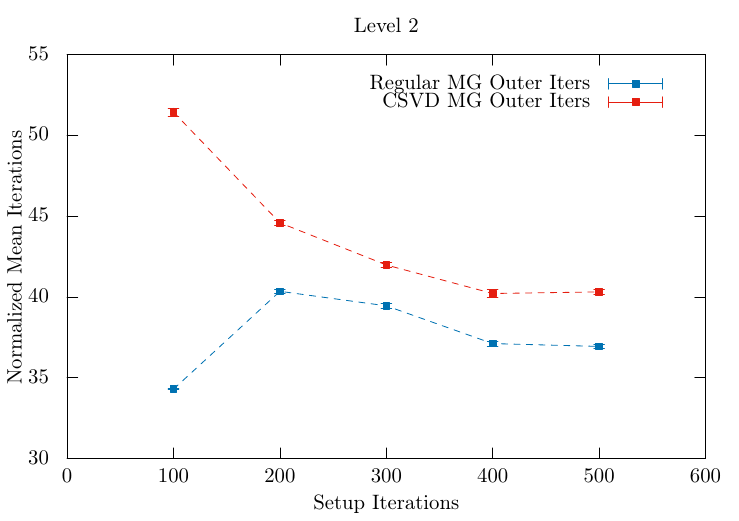}
        \includegraphics[scale=0.45]{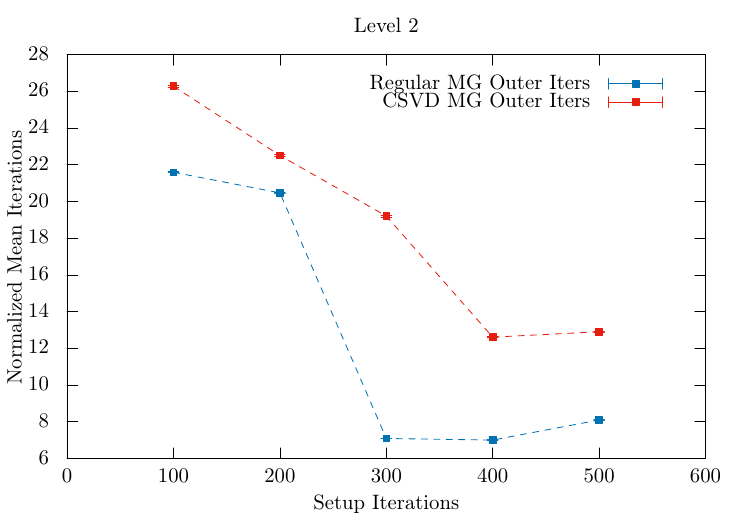}
		\caption{\label{fig::norm_iters}The normalized number of iterations on level $\ell = 1$ (top) and $\ell = 2$ (bottom) for Ensemble A (left) and Ensemble B (right).}
	\end{center}
\end{figure}

While the chiral rank-$k$ truncation setup method is observed to decrease the total cost of solving the linear equations, it is important to quantify the performance of the preconditioner on each level per iteration due to the use of the $K$-cycle. In a $K$-cycle, multiple iterations of the solver on level $\ell+1$ occur for each iteration on level $\ell$. To examine the cost per iteration on each level, the number of iterations on level $\ell+1$ is normalized by the number of iterations on level $\ell$. Fig.~\ref{fig::norm_iters} displays the normalized cost for levels $\ell = 1,2$ on both ensembles. It is observed for both ensembles across all number of setup iterations that the cost per iteration is greater for the preconditioner that utilizes the chiral rank-$k$ truncation. This is consistent with expectations that more information about the low frequency components of the error is being transferred to the coarse grids where more iterations can be performed at significantly less cost.

\subsection{Lattice Volume Scaling}
We now examine the volume dependence of the set up methods with and without the chiral rank-$k$ truncation as the volume of the hypercubic lattice is increased. The system of linear equations are solved on three lattice volumes of spatial extent $L_s = 24,32,40$ and temporal extent $L_t = 64$ and $m_{\pi} \approx$ 220 MeV with the Clover on HISQ action\footnote{The same lattice of $L_s = 32$ is used here as in the preceding numerical results.}. The parameters of the calculation were chosen from the near optimal setup parameters and used across all three volumes. 
\begin{figure}[!h]
	\begin{center}
	\includegraphics[scale=0.5]{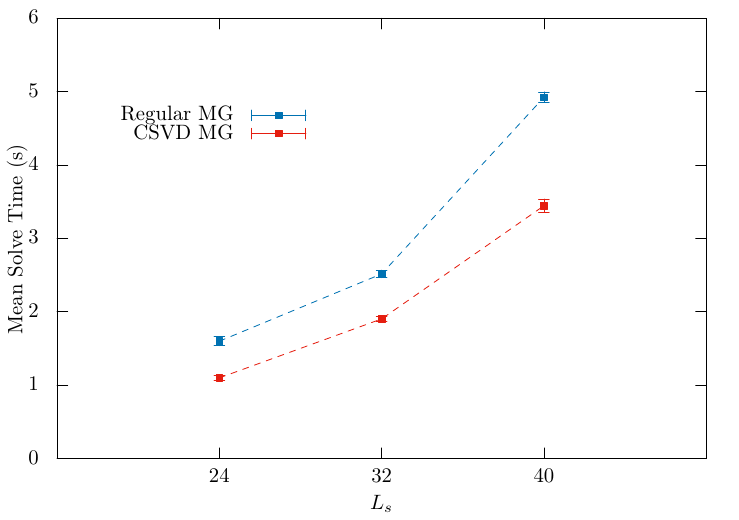}
	\label{fig::volume}
	\caption{\label{fig::volume}The performance of both MG preconditioners as the lattice volume is increased.}
	\end{center}
\end{figure}
To examine the scaling with the lattice volume, the mean execution time of the system of linear equations preconditioned by both methods is measured across all three volumes, shown in Fig.~\ref{fig::volume}. For all three volumes, the preconditioner using chiral rank-$k$ truncations shows a decrease in execution time in comparison to the conventional method. 

\section{Summary}
We have presented a modification to the setup algorithm for the multigrid
preconditioner for systems of linear equations of Wilson fermions. The chiral
rank-k truncation utilizes a singular value decomposition of the chiral components of the test vectors restricted to a domain of the lattice. By calculating
out a large basis of test vectors, the singular value decomposition is able to
truncate the basis to the fewest vectors containing the largest contribution
to the low rank approximation of the basis. In all numerical experiments, the
use of the chiral rank-k truncation results in a decrease in the time required
to solve the system of linear equations. 

\acknowledgments
\noindent TW acknowledges partial funding from the Exascale Computing Project (ECP), Project Number: 17-SC-20-SC, a collaborative effort of the U.S. Department of Energy, Office of Science and the National Nuclear Security Administration, as well as from a Royal Society Research Fellowship and the U.K. Science and Technology Facilities Council (STFC) [grant numbers ST/T000694/1, ST/X000664/1]. TW also acknowledges support from Science Foundation Ireland [grant number 21/FFP-P/10186] and Deutsche Forschungsgemeinschaft (DFG, German Research Foundation) as part of the CRC 1639 NuMeriQS – project no. 511713970. AS and ER acknowledge partial support by DOE SciDAC-5 grant (DE-FOA-0002589) and AS acknowledges partial support from National Science Foundation grant IIS-2008557. Part of this work was performed using computing resources at William \& Mary with the software codes {\tt Chroma}~\cite{chroma}, {\tt QPhiX}~\cite{qphix} and {\tt MG\_PROTO}~\cite{mgproto}. We thank the MILC Collaboration and the Hadron Spectrum Collaboration (\url{www.hadspec.org}) for making the gauge configurations available to us. The gauge configurations of the Hadron Spectrum Collaboration were generated using resources awarded from the U.S. Department of Energy INCITE program at Oak Ridge National Lab, the NSF Teragrid at the Texas Advanced Computer Center and the Pittsburgh Supercomputer Center, as well as at Jefferson Lab. We also thank Steven Gottlieb and Walter Wilcox for their help in obtaining the MILC configurations.\\


\begin{thebibliography}{99}

\bibitem[1]{wilsonmg}
J. Brannick et al., \emph{Adaptive Multigrid Algorithm for Lattice QCD},
\href{https://doi.org/10.1103/PhysRevLett.100.041601}
{\emph{Phys. Rev. Lett.} 100 (2008) 041601}
[\href{https://doi.org/10.48550/arXiv.0707.4018}{\tt 0707.4018}].

\bibitem[2]{domainwallmg}
Richard C. Brower et al., \emph{Multigrid for chiral lattice fermions: Domain wall},
\href{https://doi.org/10.1103/PhysRevD.102.094517}
{\emph{Phys. Rev. D} 102 (2020) 094517}
[\href{https://doi.org/10.48550/arXiv.2004.07732}{\tt 2004.07732}].

\bibitem[3]{staggmg}
Richard C. Brower et al., \emph{Multigrid algorithm for staggered lattice fermions},
\href{https://doi.org/10.1103/PhysRevD.97.114513}
{\emph{Phys. Rev. D} 97 (2018) 114513}
[\href{https://doi.org/10.48550/arXiv.1801.07823}{\tt 1801.07823}].

\bibitem[4]{overlapmg}
James Brannick et al., \emph{Multigrid Preconditioning for the Overlap Operator in Lattice QCD},
\href{https://doi.org/10.1007/s00211-015-0725-6}
{\emph{Numer. Math.} 132 (2016) 463}
[\href{https://doi.org/10.48550/arXiv.1410.7170}{\tt 1410.7170}].

\bibitem[5]{ddalphamg}
A. Frommer et al., \emph{Adaptive Aggregation-Based Domain Decomposition Multigrid for the Lattice Wilson--Dirac Operator},
\href{https://doi.org/10.1137/130919507}
{\emph{SIAM J. Sci. Comput.} 36 (2014) A1581}
[\href{https://doi.org/10.48550/arXiv.1303.1377}{\tt 1303.1377}].

\bibitem[6]{twistedmassmg}
S. Bacchio et al., \emph{DDalphaAMG for Twisted Mass Fermions},
\href{https://doi.org/10.22323/1.256.0259}
{\emph{PoS} \textbf{LATTICE2016} (2016) 259}
[\href{https://doi.org/10.48550/arXiv.1611.01034}{\tt 1611.01034}].

\bibitem[7]{brannick_schwinger}
J. Brannick and K. Kahl, \emph{Bootstrap Algebraic Multigrid for the 2D Wilson Dirac system},
\href{https://doi.org/10.1137/130934660}
{\emph{SIAM J. Sci. Comput.} 36 (2014) B321}
[\href{https://doi.org/10.48550/arXiv.1308.5992}{\tt 1308.5992}].

\bibitem[8]{whyte_schwinger}
Travis Whyte and Walter Wilcox and Ronald B. Morgan, \emph{Deflated GMRES with Multigrid for Lattice QCD},
\href{https://doi.org/10.1016/j.physletb.2020.135281}
{\emph{Phys. Lett. B} 803 (2020) 135281}
[\href{https://doi.org/10.48550/arXiv.1912.02868}{\tt 1912.02868}].

\bibitem[9]{morgan_twogrid}
Ronald B. Morgan et. al., \emph{Two-Grid Deflated Krylov Methods for Linear Equations},
\href{https://doi.org/10.48550/arXiv.2005.03070}{\tt 2005.03070}.

\bibitem[10]{chroma}
Robert G. Edwards and Balint Joo, \emph{The Chroma software system for lattice QCD},
\href{https://doi.org/10.1016/j.nuclphysbps.2004.11.254}
{\emph{Nucl. Phys. Proc. Suppl.} 140 (2005) 832}
[\href{https://doi.org/10.48550/arXiv.hep-lat/0409003}{\tt hep-lat/0409003}].

\bibitem[11]{qphix}
Balint Joo et al., \emph{Lattice QCD on Intel® Xeon PhiTM Coprocessors},
\href{http://dx.doi.org/10.1007/978-3-642-38750-0_4}
{\emph{Supercomputing, Lecture Notes in Computer Science}, 7905 40-54 2013 Springer Berlin Heidelberg}

\bibitem[12]{mgproto}
Balint Joo, \href{https://github.com/JeffersonLab/mg_proto/}{ \emph{MG\_PROTO: A Multigrid Library for QCD}.}

\bibitem[13]{luscher_deflation}
Martin L\"uscher, \emph{Deflation acceleration of lattice QCD simulations},
\href{https://doi.org/10.1088/1126-6708/2007/12/011}
{\emph{JHEP} \textbf{12} (2007) 011}
[\href{https://doi.org/10.48550/arXiv.0710.5417}{\tt 0710.5417}].

\bibitem[14]{luscher_coherence}
Martin L\"uscher, \emph{Local coherence and deflation of the low quark modes in lattice QCD}.
\href{https://doi.org/10.1088/1126-6708/2007/07/081}
{\emph{JHEP} \textbf{07} (2007) 081}
[\href{https://doi.org/10.48550/arXiv.0706.2298}{\tt 0706.2298}].

\bibitem[15]{isovector_charges}
Rajan Gupta et al., \emph{Isovector Charges of the Nucleon from 2+1+1-flavor Lattice QCD},
\href{https://doi.org/10.1103/PhysRevD.98.034503}
{\emph{Phys. Rev. D} 98 (2018) 034503}
[\href{https://doi.org/10.48550/arXiv.1806.09006}{\tt 1806.09006}].

\bibitem[16]{hyp}
Anna Hasenfratz and Francesco Knechtli, \emph{Flavor symmetry and the static potential with hypercubic blocking},
\href{https://doi.org/10.1103/PhysRevD.64.034504}
{\emph{Phys. Rev. D} 64 (2001) 034504}
[\href{https://doi.org/10.48550/arXiv.hep-lat/0103029}{\tt hep-lat/0103029}].

\bibitem[17]{hs1}
Robert G. Edwards and Balint Joo and Huey-Wen Lin, \emph{Tuning for Three-flavors of Anisotropic Clover Fermions with Stout-link Smearing},
\href{https://doi.org/10.1103/PhysRevD.78.054501}
{\emph{Phys. Rev. D} 78 (2008) 054501}
[\href{https://doi.org/10.48550/arXiv.0803.3960}{\tt 0803.3960}].

\bibitem[18]{hs2}
Huey-Wen Lin et al., \emph{First results from 2+1 dynamical quark flavors on an anisotropic lattice: Light-hadron spectroscopy and setting the strange-quark mass},
\href{https://doi.org/10.1103/PhysRevD.79.034502}
{\emph{Phys. Rev. D} 79 (2009) 034502}
[\href{https://doi.org/10.48550/arXiv.0810.3588}{\tt 0810.3588}].

\bibitem[19]{hs3}
David J. Wilson et al., \emph{The quark-mass dependence of elastic $\pi K$ scattering from QCD},
\href{https://doi.org/10.1103/PhysRevLett.123.042002}
{\emph{Phys. Rev. Lett.} 123 (2019) 042002}
[\href{https://doi.org/10.48550/arXiv.1904.03188}{\tt 1904.03188}].

\bibitem[20]{milc1}
A. Bazovov et al., \emph{Scaling studies of QCD with the dynamical HISQ action},
\href{https://doi.org/10.1103/PhysRevD.82.074501}
{\emph{Phys. Rev. D} 82 (2010) 074501}
[\href{https://doi.org/10.48550/arXiv.1004.0342}{\tt 1004.0342}].

\bibitem[21]{milc2}
A. Bazovov et al., \emph{Lattice QCD Ensembles with Four Flavors of Highly Improved Staggered Quarks},
\href{https://doi.org/10.1103/PhysRevD.82.074501}
{\emph{Phys. Rev. D} 87 (2013) 054505}
[\href{https://doi.org/10.48550/arXiv.1212.4768}{\tt 1212.4768}].

\bibitem[22]{chow}
Edmond Chow, \emph{An Aggregation Multilevel Method Using Smooth Error Vectors},
\href{https://doi.org/10.1137/040608192}
{\emph{SIAM J. Sci. Comput.} 27 (2006) 1727}.

\bibitem[23]{dambra}
Pasqua D'Ambra and Panayot S. Vassilevski, \emph{Improving solve time of aggregation-based adaptive AMG},
\href{https://doi.org/10.1002/nla.2269}
{\emph{Num. Lin. Alg. with App.} 26 (2019) 2269}.

\bibitem[24]{adapt}
M. Brezina et al., \emph{Adaptive Smoothed Aggregation ($\alpha$SA)},
\href{https://doi.org/10.1137/S1064827502418598}
{\emph{SIAM J. Sci. Comput.} 25 (2004) 1896}.

\bibitem[25]{kcycl}
Yvan Notay and Panayot S. Vassilevski, \emph{Recursive Krylov-based multigrid cycles},
\href{https://doi.org/10.1002/nla.542}
{\emph{Num. Lin. Alg. with App.} 15 (2008) 473}.

\bibitem[26]{stout}
Colin Morningstar and Mike Peardon, \emph{Analytic smearing of SU(3) link variables in lattice QCD},
\href{https://doi.org/10.1103/PhysRevD.69.054501}
{\emph{Phys. Rev. D} 69 (2004) 054501}
[\href{https://doi.org/10.48550/arXiv.hep-lat/0311018}{\tt hep-lat/0311018}].

\end{thebibliography}
\end{document}